\documentclass{emulateapj}
\accepted{2007 December 26}

\slugcomment{}
\shortauthors{Auger et al.}
\shorttitle{Environments of Moderate Redshift Radio Galaxies}

\newcommand{\dtable}{deluxetable*}

\begin{document}

\title{The Environments of Low and High Luminosity Radio Galaxies at Moderate Redshifts}

\author{M. W. Auger, R. H. Becker, C. D. Fassnacht}
\affil{
   Department of Physics, University of California, 1 Shields Avenue,
   Davis, CA 95616, USA }
\email{mauger@physics.ucdavis.edu}

\begin{abstract}
In the local Universe, high-power radio galaxies live in lower density environments than low-luminosity radio galaxies. If this trend continues to higher redshifts, powerful radio galaxies would serve as efficient probes of moderate redshift groups and poor clusters. Photometric studies of radio galaxies at $0.3 \lesssim z \lesssim 0.5$ suggest that the radio luminosity-environment correlation disappears at moderate redshifts, though this could be the result of foreground/background contamination affecting the photometric measures of environment. We have obtained multi-object spectroscopy of in the fields of 14 lower luminosity ($L_{\rm{1.4~GHz}} < 4\times10^{24}~\rm{W~Hz^{-1}}$) and higher luminosity ($L_{\rm{1.4~GHz}} > 1.2\times10^{25}~\rm{W~Hz^{-1}}$) radio galaxies at $z \approx 0.3$ to spectroscopically investigate the link between the environment and the radio luminosity of radio galaxies at moderate redshifts. Our results support the photometric analyses; there does not appear to be a correlation between the luminosity of a radio galaxy and its environment at moderate redshifts. Hence, radio galaxies are not efficient signposts for group environments at moderate redshifts.
\end{abstract}

\section{INTRODUCTION}
Photometric studies of radio galaxies have shown that Fanaroff-Riley type I \citep[FR I;][]{fr} (low luminosity and edge-dimmed radio lobes) and FR II (high luminosity and edge-brightened lobes) radio galaxies exist in different environments in the local Universe; FR I galaxies reside in rich clusters while FR II galaxies tend to inhabit poor clusters or rich groups \citep[e.g.,][]{heckman,prestage,lilly}. Studies of higher redshift radio galaxies indicate that this difference disappears at moderate redshifts ($z \sim 0.5$), where both FR types are associated with cluster-like environments \citep[e.g.,][]{hill,bahcall,allington}. In contrast, \citet{zirbel} suggests that, although the environments of both types of galaxies are more dense at moderate redshifts than low redshifts, the environments of FR II galaxies remain less dense than FR I sources out to $z \sim 0.5$. Only the low redshift trend of FR I environments being more dense than the environments of FR II galaxies has been confirmed with X-ray observations \citep{miller99} and optical spectroscopy \citep{miller}. It is quite difficult to detect and characterize group-like over-densities at moderate redshifts without spectroscopy. Photometric properties are difficult to use because groups tend not to have substantial bright early-type populations \citep[e.g.,][]{wilman} that form a red sequence, and foreground and background interlopers are common. Therefore, field spectroscopy is necessary to determine whether FR II galaxies may be useful signposts to identify moderate redshift galaxy groups.
 
These galaxy groups and poor clusters are excellent laboratories for studying galaxy evolution. Over half of all galaxies in the local Universe are members of groups \citep{huchra,tully}, and these groups have moderate densities and low velocity dispersions that make them the preferred environments for dynamic galaxy evolution \citep[e.g.,][]{zabludoff,carlberg,aarseth,barnes} compared to clusters or the field \citep[e.g.,][]{aceves,lin,conselice,patton}. Though cluster cores also promote some galaxy processing and evolution, the high velocity dispersions associated with these cores decrease the effective interaction cross-section between galaxies and limit the evolutionary effects of galaxy-galaxy interactions \citep[e.g.,][]{bower}. Therefore galaxy groups provide excellent conditions for studying interactions that affect the star formation rate of galaxies and AGN ignition \citep[e.g.,][]{mihos,iono,bahcallj,best}.

The most significant obstacle to studying galaxy evolution in group environments is identifying non-local groups. The Sloan Digital Sky Survey \cite[SDSS;][]{york} and the 2 Degree Field Galaxy Redshift Survey \cite[2dFGRS;][]{colless} have extended local group samples \citep[e.g.,][]{ramella,zabludoff} to redshifts $z \sim 0.2$ \citep[e.g.,][]{weinmann,collister,padilla}. Blind redshift surveys at higher redshifts have successfully identified galaxy groups to $z \sim 0.7$ \citep[e.g.,][]{wilman,gerke}, but the substantial amount of telescope time invested in these surveys makes it desirable to find alternative methods of identifying moderate redshift groups. \citet{mulchaey} use deep X-ray observations to select groups at moderate redshifts, though this method may preferentially select the most massive groups. Gravitational lenses have also been used to identify moderate redshift groups \citep{momcheva,williams,augera,augerb,fassnacht}, though the scarcity of lenses limits the utility of this method. Radio galaxies have been used extensively to find high redshift clusters \citep[e.g.,][]{deltorn,blantona,blantonb} and protoclusters \citep[e.g.,][]{miley,venemans}, but the FR I/FR II environment dichotomy suggests that they may also be useful for identifying less rich galaxy groups \citep[e.g.,][]{allington}.

In this paper we present spectroscopic observations of the fields of a sample of 14 moderate redshift radio galaxies. We investigate the relationship between the radio luminosity and spectroscopic group properties, and we analyze the effectiveness of using high luminosity radio galaxies to preferentially locate moderate redshift groups. Throughout this paper we assume a radio spectral index of $\alpha = 0.5$, where $S_{\nu} \propto \nu^{-\alpha}$, and we use a $\Lambda$CDM cosmology with $\Omega_{\Lambda} = 0.73$ and $\Omega_M = 0.27$.

\section{RADIO GALAXY SAMPLE}
We have used the SDSS and FIRST survey \citep{becker} to identify a sample of moderate redshift radio galaxies. The sources in the FIRST catalog were matched against the galaxies in the SDSS DR 5 spectroscopic catalog \citep{dr5} with a conservative matching tolerance of 1\arcsec; 99.9\% of these matches are real, as determined by comparing the number of matches from the two catalogs to the number of matches between the SDSS and a pseudo-random radio catalog (the FIRST catalog with a 2\arcmin~offset applied to all sources). The resulting matched catalog was then searched for all objects that had at least two additional FIRST sources within 60\arcsec~of the host galaxy. These sources were all reviewed by eye to morphologically identify `classical' FR I and FR II sources. Note that our radio galaxy catalog differs slightly from the usual definition of FR I and FR II sources, as all of our radio galaxies also have radio cores.

\begin{figure*}
\begin{center}
\epsscale{0.45}
 \includegraphics[width=0.45\textwidth]{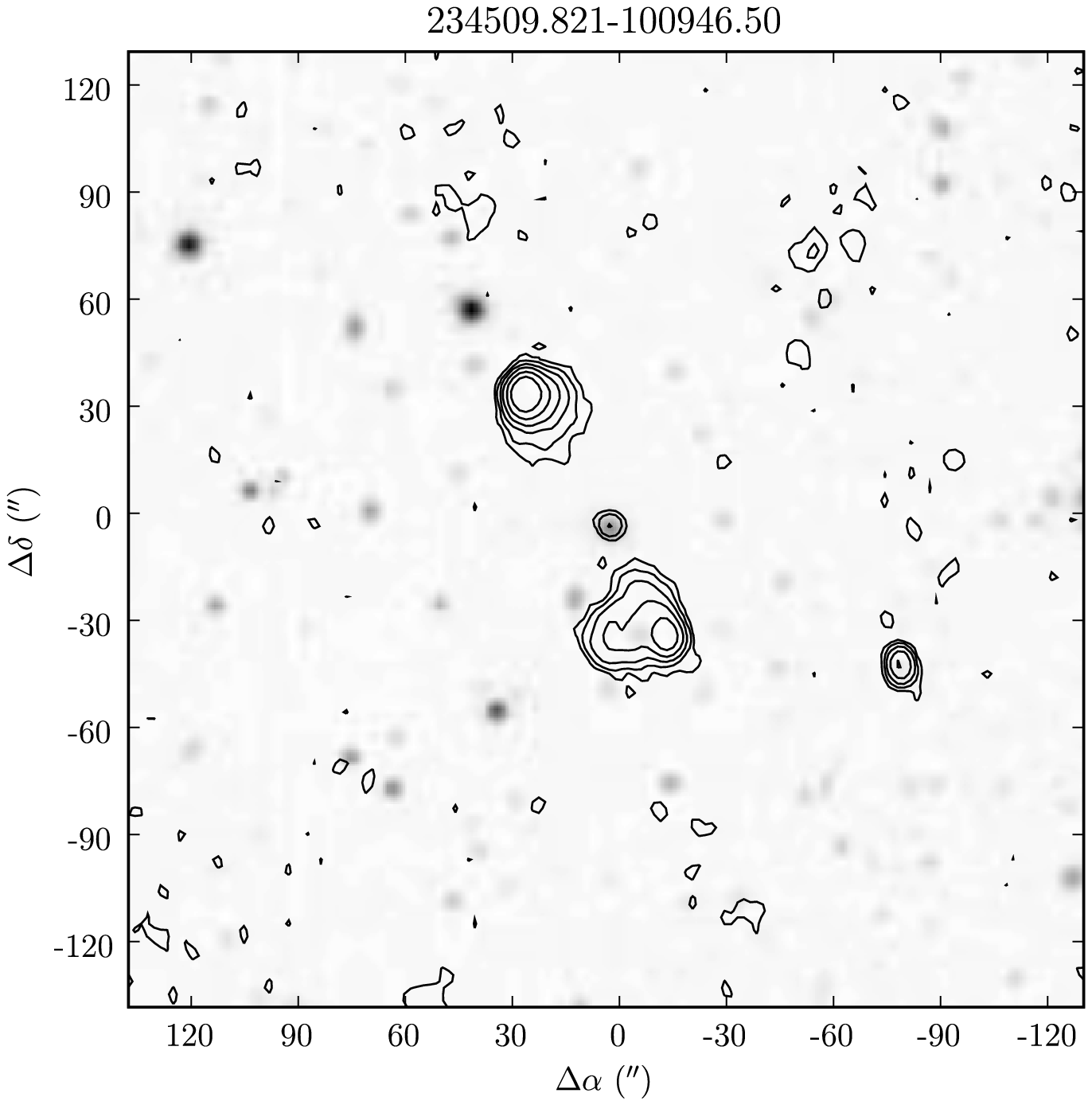}
 \includegraphics[width=0.45\textwidth]{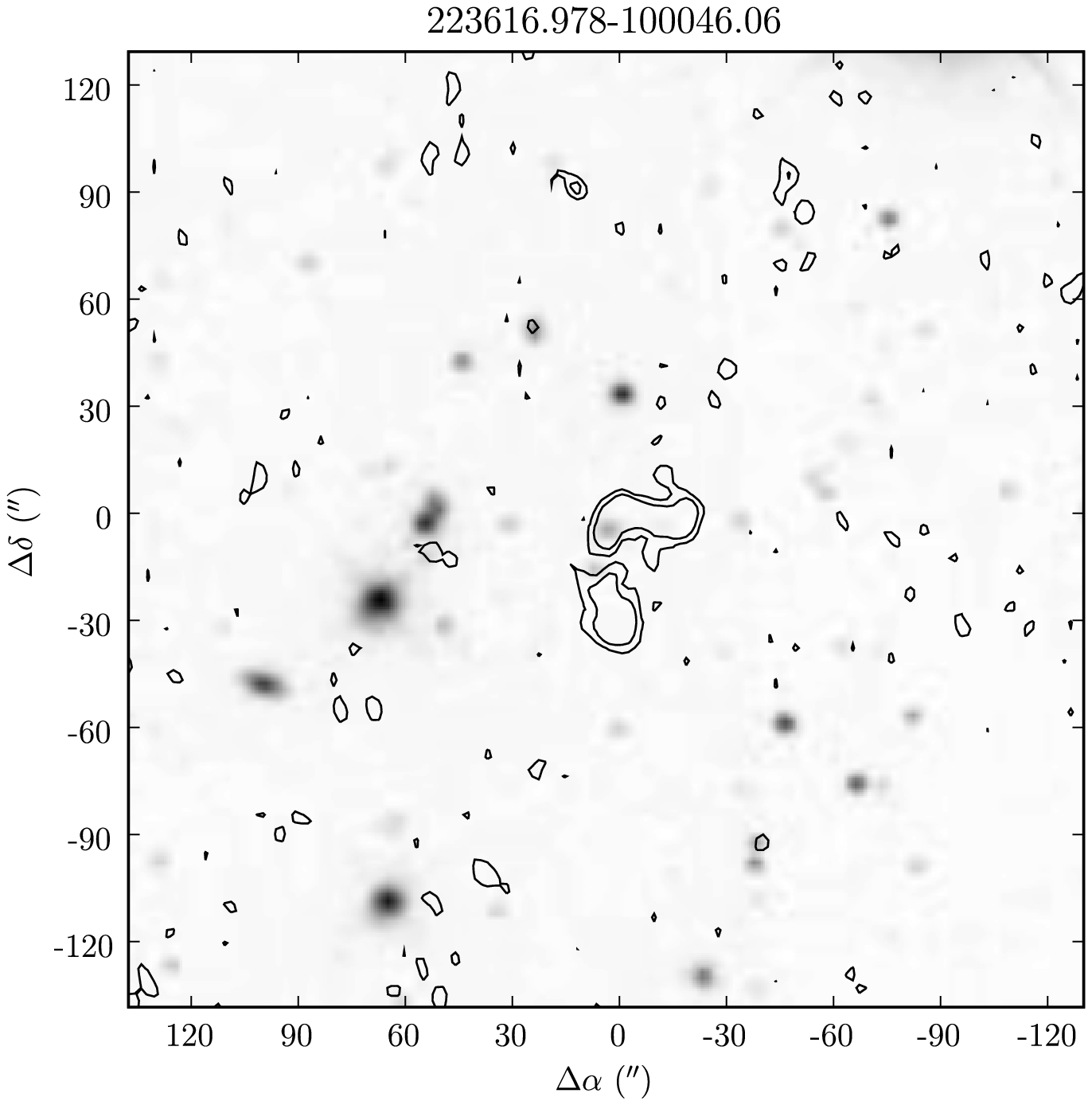}
  \caption{FIRST radio contours overlaid on the SDSS $r$ optical imaging for the radio galaxies 234509.821$-$100946.50 (a typical high-power source, \textit{left}) and 223616.932$-$100045.54 (a typical low-power source, \textit{right}). Images for all 14 of the radio galaxies are available in the electronic edition.}
\label{figure_galaxies}
\end{center}
\end{figure*}

The radio luminosities of these galaxies were determined by summing the fluxes of all detected FIRST components that are part of the radio galaxy. We also summed the fluxes of NVSS \citep{nvss} sources, and we choose the larger of the FIRST/NVSS values to assign the radio luminosity to the galaxy; this allows us to account for the sensitivity differences of the surveys and for the possibility of flux being resolved out of the FIRST images. The radio galaxies were then divided into lower luminosity ($L_{\rm{1.4~GHz}} < 4\times10^{24}~\rm{W~Hz^{-1}}$) and higher luminosity ($L_{\rm{1.4~GHz}} > 1.2\times10^{25}~\rm{W~Hz^{-1}}$) subsamples. The high luminosity criterion is motivated by previous studies of powerful radio galaxies \citep[e.g.,][]{zirbel}, while the low luminosity criterion was chosen to provide a reasonable number of sources (given the depth of the FIRST survey at $z \approx 0.35$) and to be well-separated from the high luminosity sample. Multi-object spectroscopy was obtained for the fields of the 14 targets that were at the lowest airmass during our allotted telescope time; this includes 7 low-luminosity and 7 high-luminosity sources (Table \ref{table_sources} and Figure \ref{figure_galaxies}). 

\begin{\dtable}{lllcr}
\tabletypesize{\scriptsize}
\tablecolumns{5}
\tablewidth{0pc}
\tablecaption{Moderate Redshift Radio Galaxy Sub-Sample}
\tablehead{
  \colhead{} & \colhead{} & \colhead{} & \colhead{} &
  \colhead{$L_{\rm{1.4 GHz}}$} \\
  \colhead{Radio Galaxy} &
  \colhead{R.~A.} &
  \colhead{Dec} &
  \colhead{Redshift} &
  \colhead{$10^{24}~\rm{W~Hz^{-1}}$}
}
\startdata
125622.106$+$012535.08  &  12 56 22.11  &  $+$01 25 35.1  &  0.3346  &  22.86  \\
141652.862$+$120227.14  &  14 16 52.86  &  $+$12 02 27.1  &  0.3344  &  3.41  \\
144020.784$+$112507.09  &  14 40 20.78  &  $+$11 25 07.1  &  0.3109  &  20.50  \\
155437.010$+$374406.03  &  15 54 37.01  &  $+$37 44 06.0  &  0.3956  &  22.70  \\
165059.834$+$190110.02  &  16 50 59.83  &  $+$19 01 10.0  &  0.3203  &  3.88  \\
170653.810$+$230903.52  &  17 06 53.81  &  $+$23 09 03.5  &  0.3295  &  3.53  \\
213739.115$-$081403.44  &  21 37 39.12  &  $-$08 14 03.4  &  0.3416  &  1.91  \\
222821.192$+$011412.29  &  22 28 21.19  &  $+$01 14 12.3  &  0.2934  &  2.09  \\
223616.932$-$100045.54  &  22 36 16.93  &  $-$10 00 45.5  &  0.3296  &  2.95  \\
232844.520$+$000134.32  &  23 28 44.52  &  $+$00 01 34.3  &  0.2921  &  3.59  \\
234509.821$-$100946.50  &  23 45 09.82  &  $-$10 09 46.5  &  0.2923  &  12.90  \\
011425.594$+$002932.59  &  01 14 25.59  &  $+$00 29 32.6  &  0.3546  &  17.01  \\
013352.730$+$011343.63  &  01 33 52.73  &  $+$01 13 43.6  &  0.3081  &  14.39  \\
021437.167$+$004235.36  &  02 14 37.17  &  $+$00 42 35.4  &  0.2902  &  15.32  \\
\enddata
\label{table_sources}
\end{\dtable}

\section{FIELD SPECTROSCOPY}
We used the SDSS $g, r, i, z$ imaging to choose targets for our multi-object spectroscopy. Empirical color distributions at the redshift of each radio galaxy were determined using the SDSS spectroscopic database \cite[e.g.,][]{augerslacs}. All of the galaxies in the field of the radio galaxy that did not already have a redshift from SDSS were given a priority based on Gaussian weights from the empirical color distributions,
$$\rm{weight} \propto e^{\frac{-\Delta^{2}_{c}}{2 \sigma^{2}_{c}}},$$
where $\Delta_{c}$ is the difference between the observed color and the mean of the color distribution and $\sigma_{c}$ is the quadrature sum of the width of the color distribution and the errors on the SDSS photometry. Higher priority was given to galaxies with $r$ less than one magnitude brighter and 2.5 magnitudes fainter than the radio galaxy. Slitmasks were made to optimize the number of high priority objects and the remaining space on the slitmasks was filled using the lower-priority objects.

Our spectroscopy of the fields of the radio galaxies was obtained with the Low-Resolution Imaging Spectrometer \citep[LRIS;][]{oke} on the Keck I telescope (Table \ref{table_spectroscopy}). All of the observations employed the 300/5000 grism on the blue side with dispersion $\sim 1.41 \rm{\AA~pix}^{-1}$, the 600/10000 grating with dispersion $\sim 1.25 \rm{\AA~pix}^{-1}$ on the red side, and the 680 dichroic to split the beam. The conditions were typically clear with seeing $\sim 0\farcs7$ for all of the observations. The multi-slit data were reduced using an automated pipeline written in Python (Auger, {\em in prep.}). This pipeline removes the bias and overscan regions of the CCDs and flatfields the data. It then automatically determines a mask distortion and wavelength solution using the flatfield images, arclamp exposures, and night skylines. The spectra are resampled and individual exposures are coadded, then spectral traces are extracted using a flux-weighted aperture. Redshifts were determined by cross-correlating the extracted spectra with galactic and stellar template spectra, and all of these redshifts were verified by eye. If the cross-correlation failed, redshifts were determined by manually identifying features; the typical errors from the cross-correlation and manual indentification are $\Delta z \sim 0.0003$. Manual identification was necessary for only $\sim 8\%$ of the objects, most of which are at $z > 0.7$. These objects tend to require manual redshifts for two reasons: the correlation templates are redshifted to wavelengths where artifacts from skylines can confuse the cross-correlation, and the templates do not extend sufficiently blueward to cover the Mg and Fe absorption features sometimes seen in these spectra. The number of successful redshifts obtained per mask is listed in Table \ref{table_spectroscopy}. These redshifts were added to the galaxies with SDSS redshifts in each field to create redshift catalogs for the fields of the 14 radio galaxies.

\begin{deluxetable}{llcc}
\tabletypesize{\scriptsize}
\tablecolumns{4}
\tablewidth{0pc}
\tablecaption{LRIS MOS Observations}
\tablehead{
 \colhead{} &
 \colhead{Observation} &
 \colhead{Exposure} &
 \colhead{\# of} \\
 \colhead{Radio Galaxy} &
 \colhead{Date} &
 \colhead{Time (s)} &
 \colhead{Redshifts}
}
\startdata
125622.106$+$012535.08  &  2007 Jun 11  &  3600  &  26  \\
141652.862$+$120227.14  &  2007 Jun 11  &  3600  &  30  \\
144020.784$+$112507.09  &  2007 Jun 11  &  3600  &  30  \\
155437.010$+$374406.03  &  2007 Jun 11  &  3600  &  24  \\
165059.834$+$190110.02  &  2007 Jun 11  &  3000  &  27  \\
170653.810$+$230903.52  &  2007 Jun 11  &  3000  &  27  \\
213739.115$-$081403.44  &  2007 Jul 14  &  2400  &  27  \\
222821.192$+$011412.29  &  2007 Jul 15  &  2400  &  22  \\
223616.978$-$100046.06  &  2007 Jun 11  &  2700  &  22  \\
                      &  2007 Jul 14  &  2400  &  19  \\
232844.518$+$000133.64  &  2007 Jun 12  &  3600  &  22  \\
                      &  2007 Jul 15  &  2400  &  20  \\
234509.821$-$100946.50  &  2007 Jul 14  &  2400  &  22  \\
011425.594$+$002932.59  &  2007 Jul 14  &  2400  &  22  \\
013352.730$+$011343.63  &  2007 Jul 15  &  2400  &  26  \\
021437.167$+$004235.36  &  2007 Jul 15  &  2400  &  23  \\
\enddata
\label{table_spectroscopy}
\end{deluxetable}

\section{ANALYSIS AND RESULTS}
We use the iterative group finding algorithm of \citet{augera} to determine group membership for each of the radio galaxy groups. The algorithm requires an initial estimate of the velocity dispersion, which we take to be a typical poor cluster dispersion of $700~\rm{km~s^{-1}}$, and defines the group membership by requiring that all group galaxies be clustered spatially and kinematically. The iterative method determines group velocity dispersions using the gapper algorithm when the group has fewer than 15 members and the biweight statistic when the group has 15 or more members \citep[e.g.,][]{beers}, and the group center is the mean position of the group galaxies. The errors on the velocity dispersions are determined from a bootstrap resampling and the offset to the radio galaxy is with respect to the centroid of the group. The group properties of the 14 radio galaxies are listed in Table \ref{table_groups}, and the properties of the members of each group are tabulated in Table \ref{table_groupgals}.

\begin{\dtable}{lllrrrrcr}
\tabletypesize{\scriptsize}
\tablecolumns{9}
\tablewidth{0pc}
\tablecaption{Radio Galaxy Group Properties}
\tablehead{
 \colhead{}  &
 \colhead{}  &
 \colhead{}  &
 \colhead{}  &
 \colhead{}  &
 \colhead{}  &
 \colhead{Offset}  &
 \colhead{}  &
 \colhead{$\sigma_{group}$}  \\
 \colhead{Name} &
 \colhead{R.~A.} &
 \colhead{Dec} &
 \colhead{N$_{spec}$} &
 \colhead{N$_{500}$} &
 \colhead{N$_{1000}$} &
 \colhead{($^{\prime\prime}$)} &
 \colhead{$z$} &
 \colhead{($\rm{km~s^{-1}}$)} 
}
\startdata
125622.106$+$012535.08  &  12 56 21.56  &  $+$01 25 46.4  &  21  &  17  &  30  &   14  &  0.3351  &  $ 390 \pm 110$  \\
141652.862$+$120227.14  &  14 16 53.33  &  $+$12 02 14.9  &  12  &   7  &  18  &   14  &  0.3340  &  $ 400 \pm  60$  \\
144020.784$+$112507.09  &  14 40 20.62  &  $+$11 25 15.7  &  16  &  15  &  33  &    9  &  0.3132  &  $ 710 \pm  100$  \\
155437.010$+$374406.03  &  15 54 36.83  &  $+$37 45 39.6  &  10  &   9  &  24  &   94  &  0.3944  &  $ 510 \pm 200$  \\
165059.834$+$190110.02  &  16 50 59.78  &  $+$19 01 46.8  &  10  &  11  &  23  &   37  &  0.3234  &  $ 580 \pm 160$  \\
170653.810$+$230903.52  &  17 06 52.51  &  $+$23 08 29.7  &  12  &  12  &  30  &   38  &  0.3286  &  $ 320 \pm  60$  \\
213739.115$-$081403.44  &  21 37 37.78  &  $-$08 14 23.0  &  25  &  26  &  45  &   28  &  0.3413  &  $ 640 \pm  90$  \\
222821.192$+$011412.29  &  22 28 22.21  &  $+$01 13 50.0  &   4  &   5  &  13  &   27  &  0.2951  &  $ 930 \pm 320$  \\
223616.932$-$100045.54  &  22 36 17.16  &  $-$10 00 17.8  &   7  &   6  &  18  &   28  &  0.3303  &  $ 530 \pm  100$  \\
232844.520$+$000134.32  &  23 28 45.39  &  $+$00 01 01.8  &   8  &   8  &  21  &   35  &  0.2940  &  $ 640 \pm 210$  \\
234509.821$-$100946.50  &  23 45 02.71  &  $-$10 11 02.6  &   8  &  11  &  34  &  130  &  0.2903  &  $ 650 \pm 150$  \\
011425.594$+$002932.59  &  01 14 25.40  &  $+$00 29 19.9  &  15  &  14  &  28  &   13  &  0.3534  &  $ 440 \pm 200$  \\
013352.730$+$011343.63  &  01 33 53.37  &  $+$01 13 38.6  &   4  &   7  &  18  &   11  &  0.3075  &  $ 230 \pm  70$  \\
021437.167$+$004235.36  &  02 14 37.84  &  $+$00 43 10.9  &   9  &  11  &  23  &   37  &  0.2894  &  $ 140 \pm  40$  
\enddata
\tablecomments{Col. (1): Name of radio galaxy. Col. (2): Right Ascension in hours, minutes, and second. Col. (3): Declination in degrees, arcminutes, and arcseconds. Col. (4): Number of spectroscopically confirmed group members. Col. (5): Number of photometrically estimated group members within 500 h$^{-1}$~kpc. Col. (6): Number of photometrically estimated group members within 1 h$^{-1}$~Mpc. Col. (7): Offset of the radio galaxy from the spectroscopic group center. Col. (8): Redshift of the group. Col. (9): Velocity dispersion of the group.} 
\label{table_groups}
\end{\dtable}

\begin{\dtable}{lllrrrrr}
\tabletypesize{\scriptsize}
\tablecolumns{8}
\tablewidth{0pc}
\tablecaption{Group Member Properties}
\tablehead{
 \colhead{Name} &
 \colhead{R.~A.}  &
 \colhead{Dec}  &
 \colhead{Redshift}  &
 \colhead{$g$}  &
 \colhead{$r$}  &
 \colhead{$i$}  &
 \colhead{$z$}
}
\startdata
125622.106+012535.08    &  12 56 30.804  &  01 25 04.836  &  0.3327  &  21.906  &  20.665  &  19.951  &  19.700  \\
    &  12 56 16.982  &  01 25 30.720  &  0.3329  &  22.788  &  20.461  &  19.831  &  19.380  \\
    &  12 56 30.506  &  01 26 02.724  &  0.3332  &  20.955  &  19.477  &  18.942  &  18.665  \\
    &  12 56 14.794  &  01 25 44.256  &  0.3335  &  22.096  &  20.519  &  19.889  &  19.535  \\
    &  12 56 27.386  &  01 26 44.016  &  0.3338  &  22.446  &  20.671  &  20.053  &  19.558  \\
    &  12 56 23.198  &  01 25 05.916  &  0.3338  &  21.727  &  20.295  &  19.667  &  19.337  \\
    &  12 56 25.534  &  01 26 26.808  &  0.3339  &  22.385  &  20.604  &  20.056  &  19.711  \\
    &  12 55 54.360  &  01 23 27.994  &  0.3340  &  20.082  &  18.378  &  17.795  &  17.384  \\
    &  12 56 20.503  &  01 26 52.620  &  0.3341  &  22.822  &  21.195  &  20.606  &  20.234  \\
    &  12 56 22.106  &  01 25 35.148  &  0.3346  &  19.711  &  17.880  &  17.261  &  16.895  \\
    &  12 56 13.978  &  01 25 04.908  &  0.3346  &  20.498  &  18.804  &  18.198  &  17.882  \\
    &  12 56 15.804  &  01 26 51.288  &  0.3347  &  21.017  &  19.401  &  18.751  &  18.442  \\
    &  12 56 26.422  &  01 26 05.244  &  0.3349  &  22.913  &  21.571  &  21.087  &  20.769  \\
    &  12 56 18.802  &  01 25 23.160  &  0.3351  &  22.184  &  20.485  &  19.896  &  19.588  \\
    &  12 56 16.642  &  01 26 30.444  &  0.3354  &  21.866  &  20.268  &  19.621  &  19.356  \\
    &  12 56 36.026  &  01 26 33.072  &  0.3358  &  21.852  &  20.592  &  20.196  &  19.940  \\
    &  12 56 34.111  &  01 26 22.236  &  0.3360  &  22.360  &  21.300  &  20.704  &  20.520  \\
    &  12 56 24.034  &  01 25 14.304  &  0.3377  &  22.575  &  21.073  &  20.371  &  19.888  \\
    &  12 56 21.091  &  01 25 01.272  &  0.3378  &  21.853  &  20.517  &  19.913  &  19.496  \\
    &  12 56 27.929  &  01 25 57.396  &  0.3392  &  20.839  &  19.061  &  18.517  &  18.189  \\
    &  12 56 11.810  &  01 25 36.156  &  0.3402  &  21.761  &  20.067  &  19.455  &  18.947  \\
\enddata
\tablecomments{Group member properties for the group associated with the radio galaxy 125622.106+012535.08. Group member properties for all 14 systems are available in the electronic edition. Col. (1): Name of radio galaxy. Col. (2): Right Ascension in hours, minutes, and seconds. Col. (3): Declination in degrees, arcminutes, and arcseconds. Col. (4): Galaxy redshift. Cols. (5-8): SDSS $g, r, i, z$ magnitudes.}
\label{table_groupgals}
\end{\dtable}

\begin{figure*} 
\begin{center}
\epsscale{0.45}
 \includegraphics[width=0.45\textwidth]{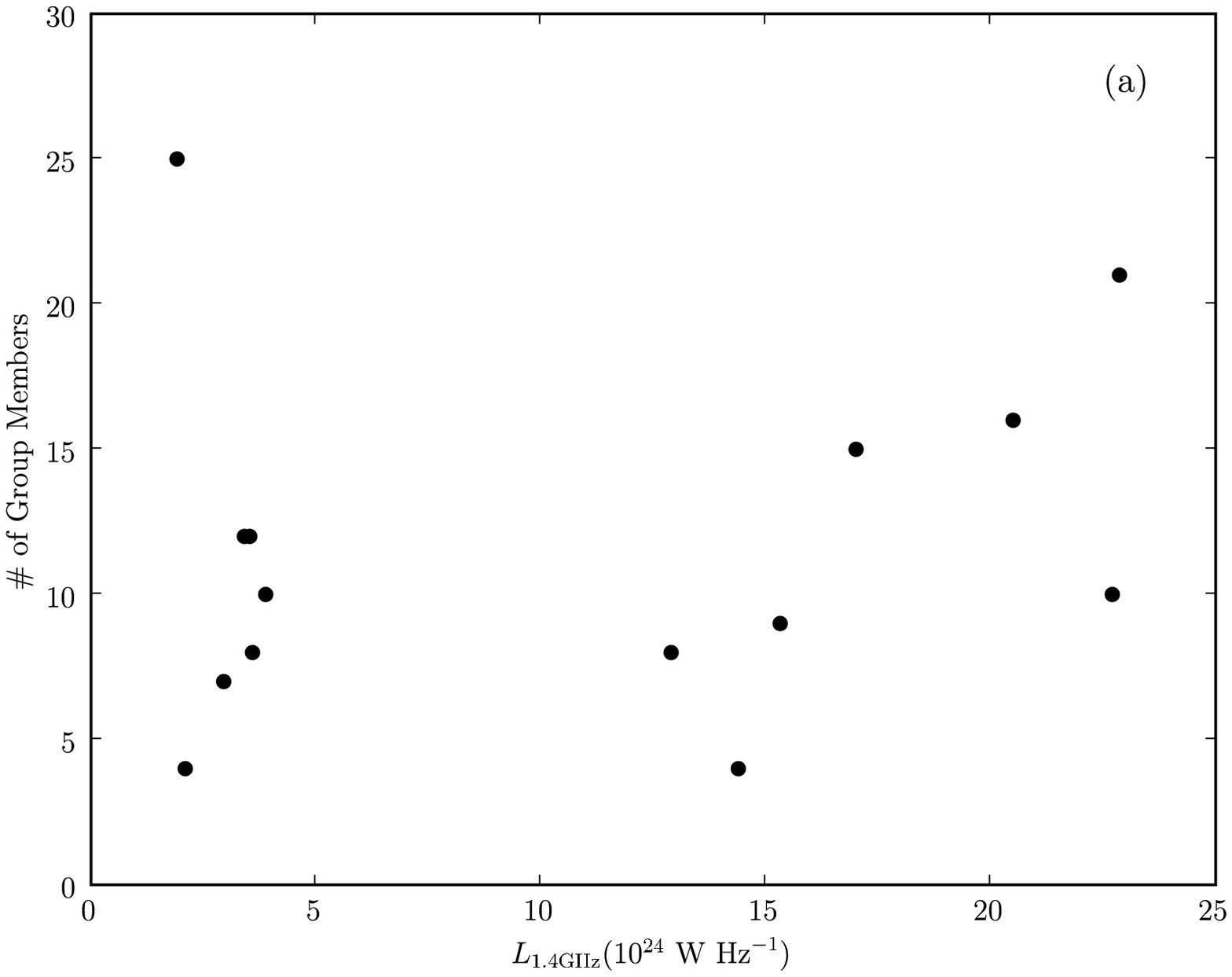}
 \includegraphics[width=0.45\textwidth]{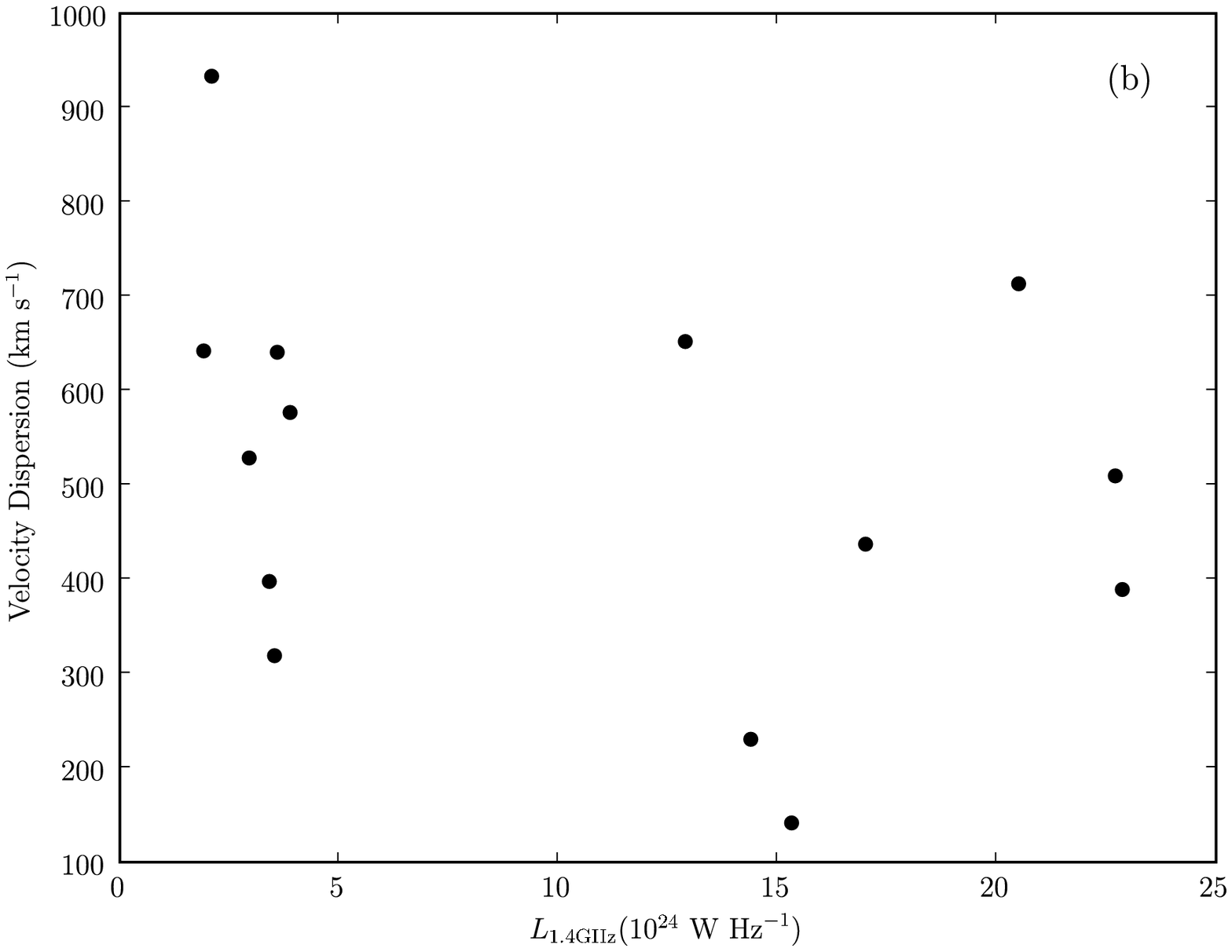}
 \includegraphics[width=0.45\textwidth]{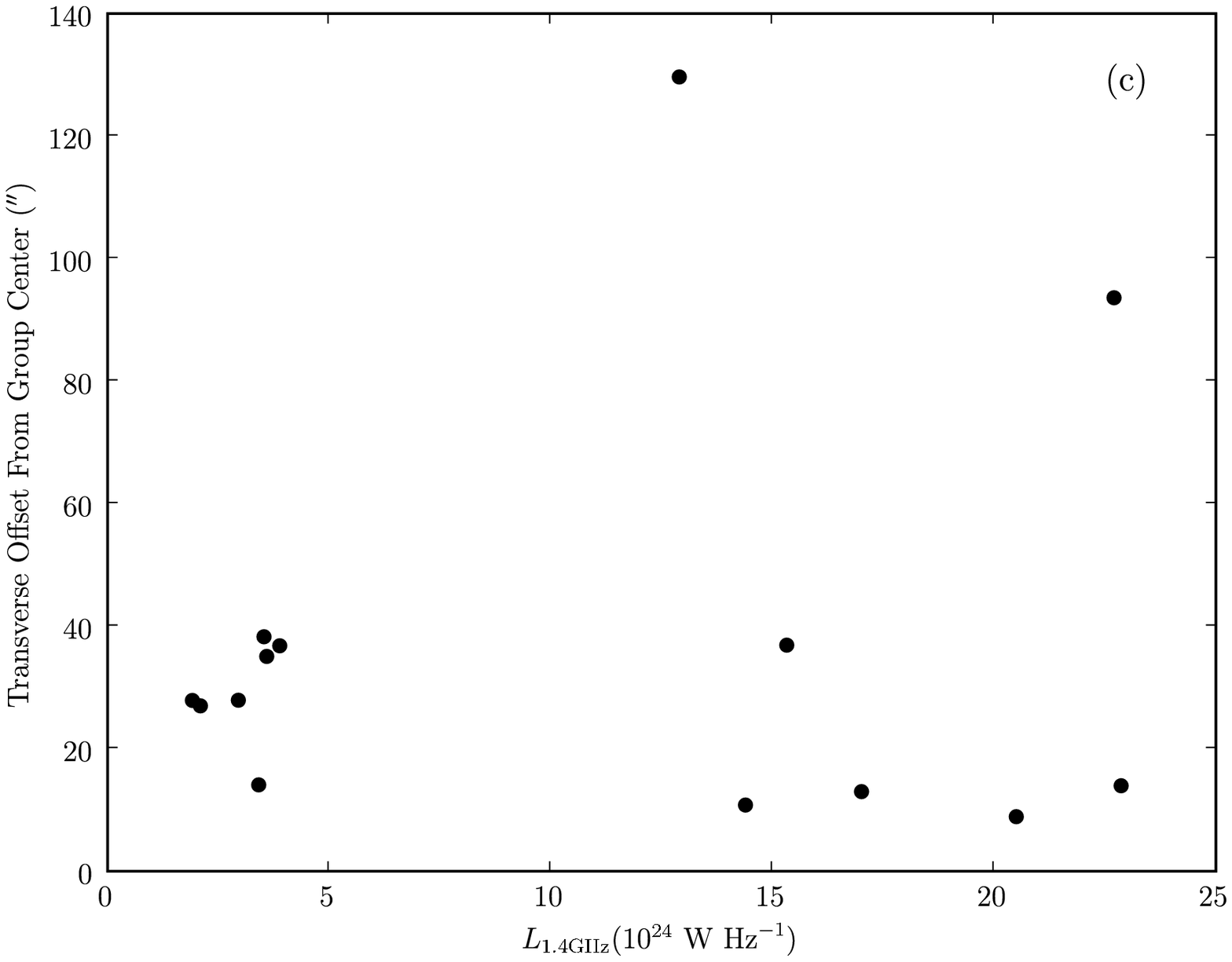}
 \includegraphics[width=0.45\textwidth]{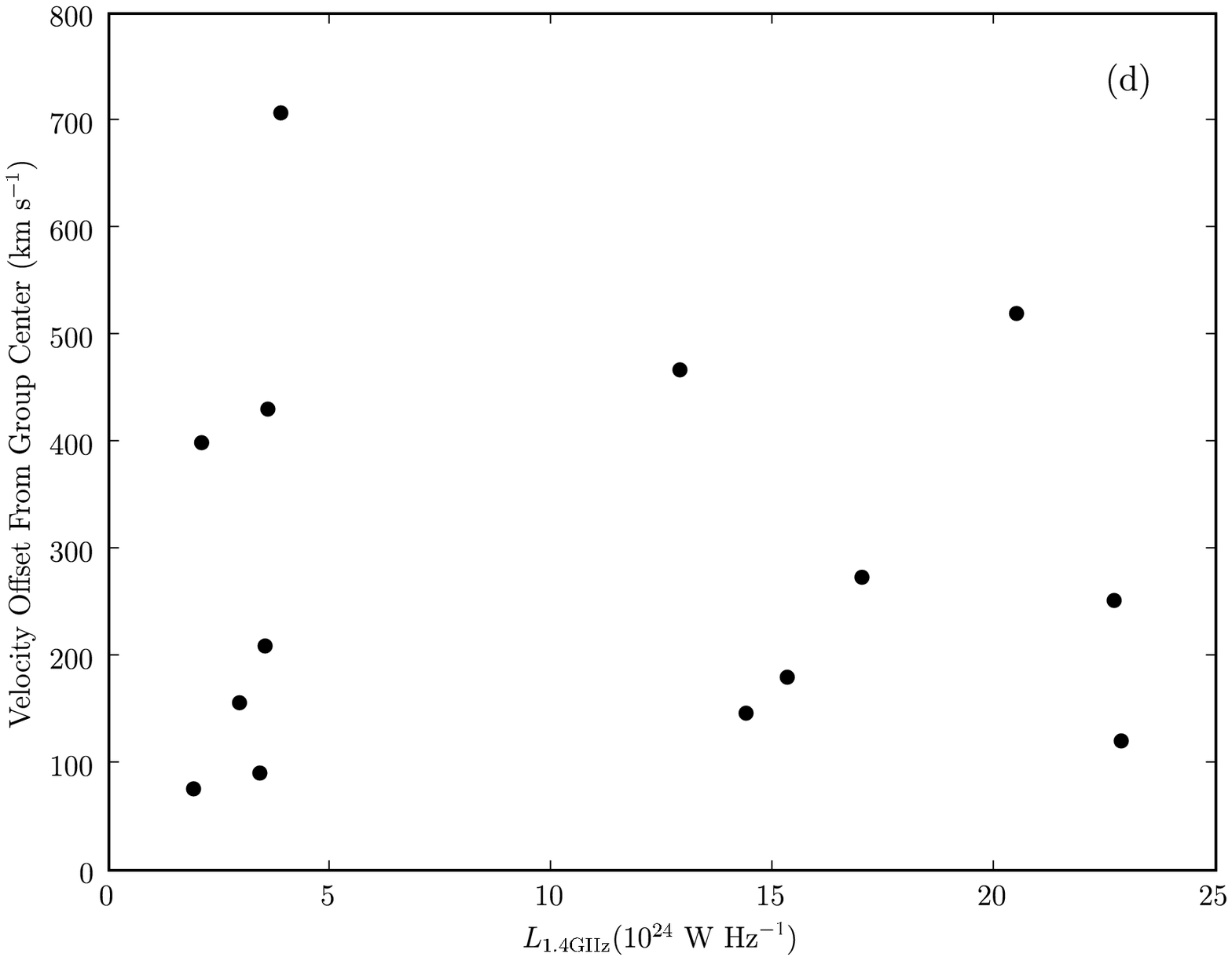}
 \includegraphics[width=0.45\textwidth]{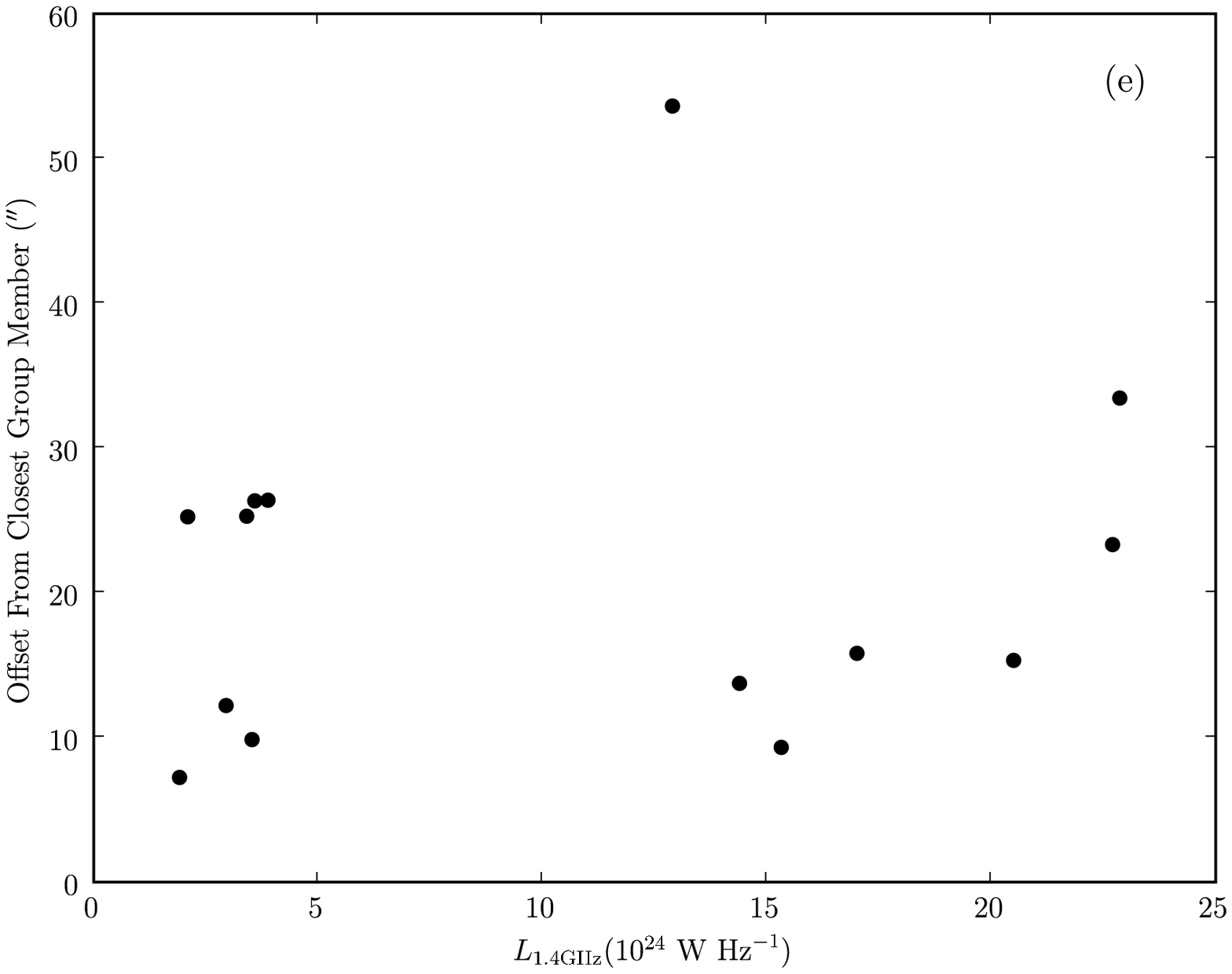}
  \caption{Relation between the radio galaxy luminosity and various group properties, including: (a) number of group members; (b) velocity dispersion of the group; (c) transverse offset of the radio galaxy from the group center; (d) kinematic offset of the radio galaxy from the group center; and (e) the distance from the radio galaxy to the closest group member.}
\label{figure_properties}
\end{center}
\end{figure*}

We have investigated several correlations between the radio luminosity and the group characteristics, including: the number of spectroscopically confirmed group members (Figure \ref{figure_properties}a); the group velocity dispersion (Figure \ref{figure_properties}b); the radio galaxy transverse offset from the group center (Figure \ref{figure_properties}c); the radio galaxy kinematic offset from the group center (Figure \ref{figure_properties}d); and the distance from the radio galaxy to the closest group member (Figure \ref{figure_properties}e). All of these properties are consistent with being uncorrelated with the radio luminosity. Furthermore, a Kolmogorov-Smirnov test cannot distinguish between the low-luminosity and high-luminosity subsamples for any of these properties.

\begin{figure*}
\begin{center}
\epsscale{0.45}
 \includegraphics[width=0.45\textwidth]{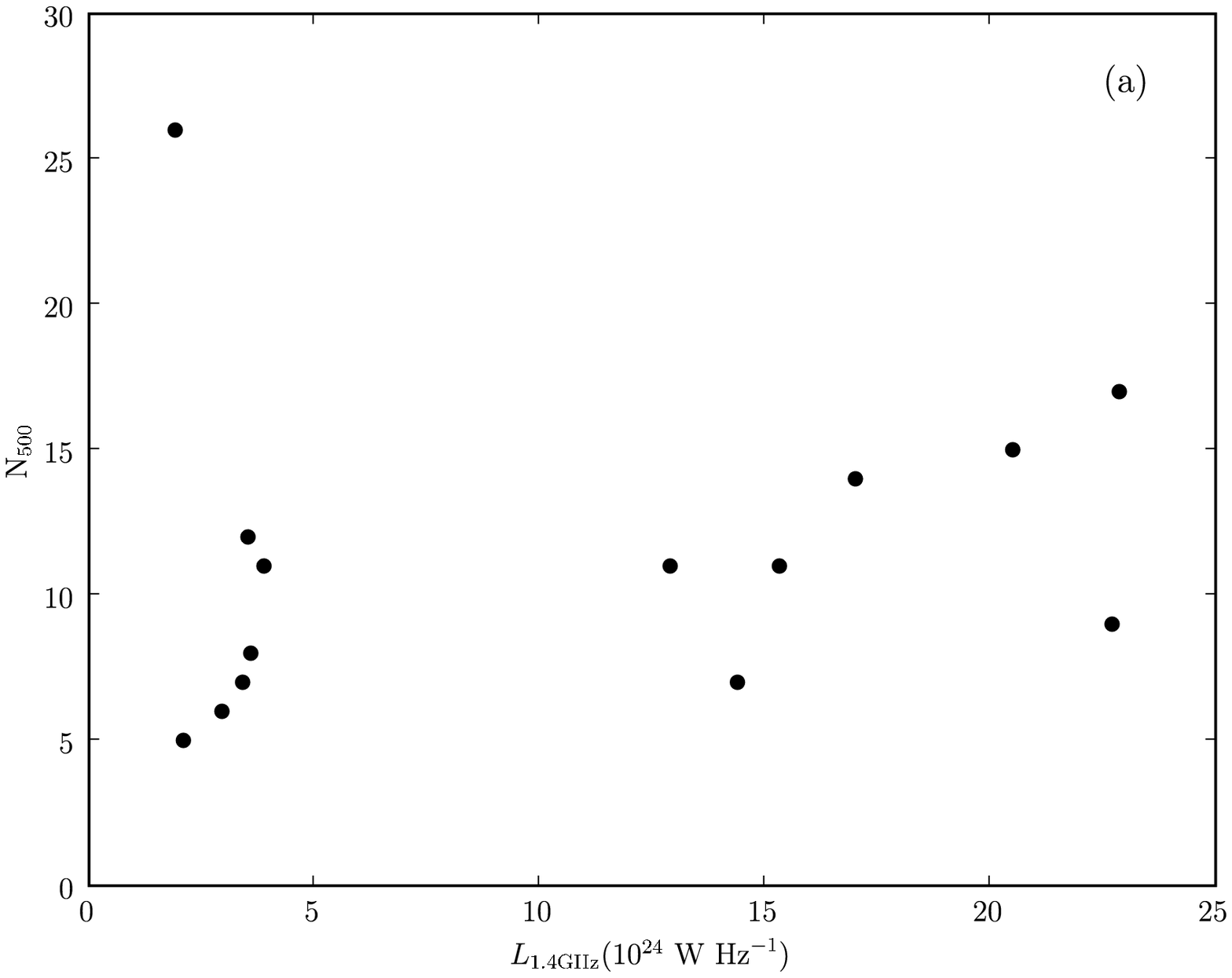}
 \includegraphics[width=0.45\textwidth]{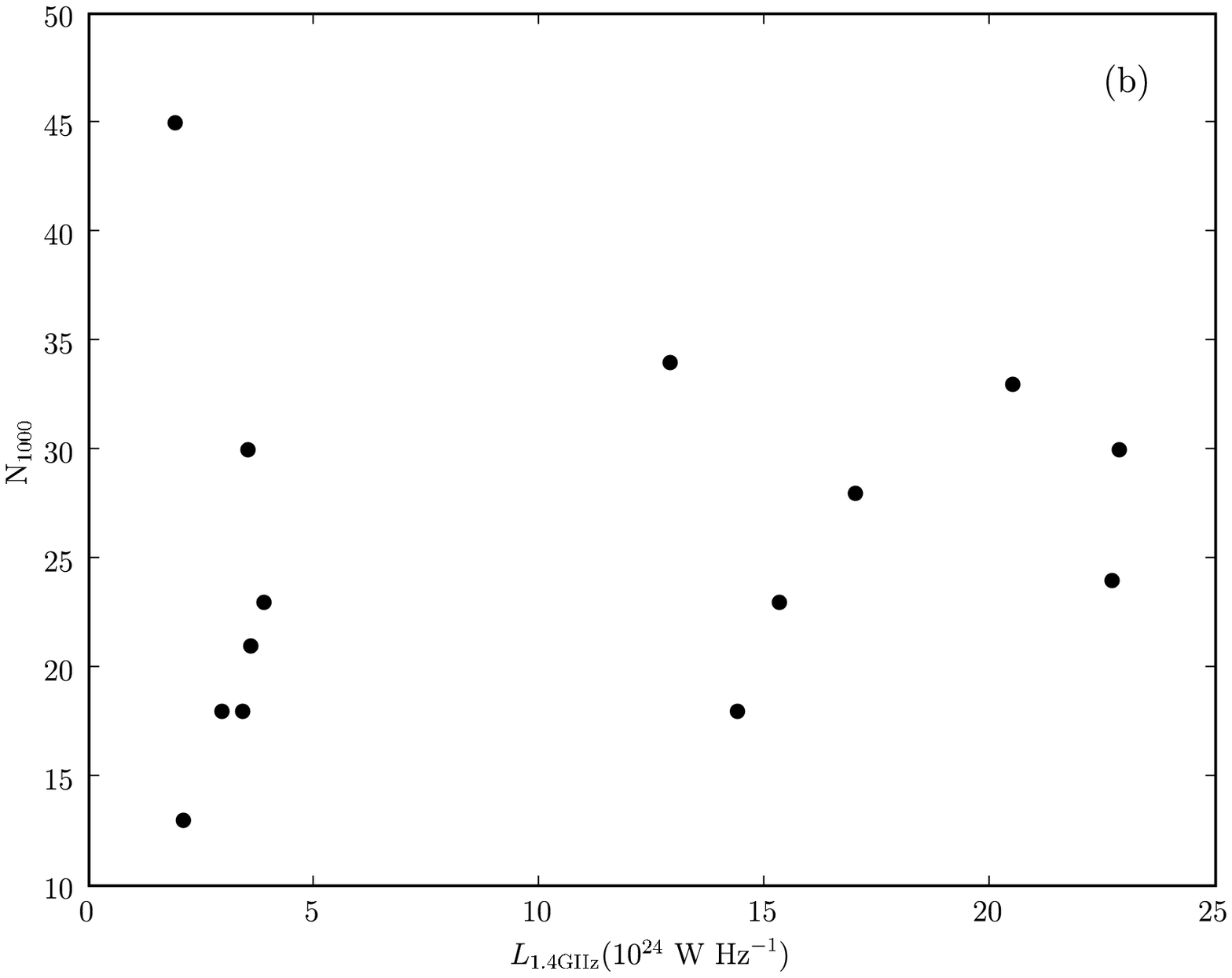}
  \caption{Relation between the radio galaxy luminosity and a photometric estimate of the number of group members associated with the radio galaxy within (a) 500~h$^{-1}$~kpc and (b) 1~h$^{-1}$~Mpc.}
\label{figure_photo}
\end{center}
\end{figure*}

Our spectroscopic completeness varies substantially across the 14 fields as a result of only obtaining one slitmask in each field. We therefore use the SDSS photometry to also characterize the radio galaxy environments. We use the priorities assigned to our spectroscopic targets to create a probability distribution that describes the likelihood that a galaxy with a given priority is spectroscopically found to be at the same redshift as the radio galaxy. We fit a smooth analytic form to the empirical distribution and assign probabilities to all of the galaxies in the field of each radio galaxy based upon their previously determined spectroscopic priority. We sum the probabilities of all galaxies within 500 h$^{-1}$~kpc and 1 h$^{-1}$~Mpc of the radio galaxies to determine the expected number of galaxies at the redshifts of the radio galaxies, N$_{500}$ and N$_{1000}$. These values are collected in Table \ref{table_groups} and provide further evidence that the radio galaxy luminosity is not correlated with the density of galaxies (Figure \ref{figure_photo}).

We have also investigated the relationship between other radio properties of the radio galaxy and the environment. The core-to-lobe flux ratio and the angular extent of the radio emission were computed using the sources detected in FIRST. We note that the FIRST data may not include resolved-out extended emission and the lobe fluxes may therefore be underestimated. However, we do not find any significant correlation between the properties of the environment and the core-to-lobe flux ratio or the angular extent of the radio galaxy.

\section{DISCUSSION}
\citet{geach} have examined the environments of a set of four radio galaxies at redshifts $z \sim 0.35$ and $z \approx 0.65$ with spectroscopy and X-ray observations. This sample includes three galaxies that we would classify as low-power and one higher-power radio galaxy, and the sample spans a range of environments. We note, however, that none of the galaxies in our sample or the \citet{geach} sample are truly isolated; all have at least two spectroscopic companions. This could be due to selection effects. \citet{geach} investigated radio galaxies that appeared to have a red sequence, while our sample required the galaxy to be in the SDSS spectroscopic sample. The SDSS preferentially observes more luminous galaxies at higher redshifts, and these galaxies tend to exist in over-dense environments. However, photometric surveys of purely radio-selected galaxies have also suggested that nearly all radio-galaxies reside in groups \citep[e.g.,][]{allington}, and we therefore do not expect this bias to be significant.

Our results spectroscopically support the results from photometric surveys of radio galaxies; low-luminosity and high-luminosity sources tend to exist in the same environments at moderate redshifts. The dichotomy in environments at low redshifts is likely due to the density of the intergalactic medium (IGM); high luminosity galaxies tend to have large extended lobes that do not form if the IGM is too dense \citep[e.g.,][]{prestage}. The absence of this dichotomy at higher redshift is probably results from cosmological evolution of the density of the IGM \citep[e.g.,][]{ettori}.

There is a substantial amount of scatter in the properties of the radio galaxy environments, though the typical radio galaxy is in a group with approximately 12 members and a velocity dispersion of $\sim 500~\rm{km~s^{-1}}$. The systems 222821.192$+$011412.29 and 013352.730$+$011343.63 are found to have only 3 companions in addition to the radio galaxy. These systems may not belong to bound groups but rather may be embedded in filaments; the large velocity spread for the low-luminosity system 222821.192$+$011412.29 further supports this conclusion. We therefore conclude that at moderate redshifts, radio galaxies reside in all types of environments, from filaments to clusters, without regard to the radio luminosity. \citet{belsole} reach a similar conclusion from a study of the X-ray environments of higher redshift radio galaxies and quasars. Thus, while radio galaxies are shown to trace large-scale-structure, they are not useful for targeting environments with specific properties (e.g., lower velocity dispersion or less rich environments) at higher redshifts.


\acknowledgments 
We would like to thank Lori Lubin for helpful comments. Many of the data presented herein were obtained at the W.M. Keck Observatory, which is operated as a scientific partnership among the California Institute of Technology, the University of California and the National Aeronautics and Space Administration. The Observatory was made possible by the generous financial support of the W.M. Keck Foundation. The authors wish to recognize and acknowledge the very significant cultural role and reverence that the summit of Mauna Kea has always had within the indigenous Hawaiian community.  We are most fortunate to have the opportunity to conduct observations from this mountain. This work has made extensive use of the SDSS database. Funding for the SDSS and SDSS-II has been provided by the Alfred P. Sloan Foundation, the Participating Institutions, the National Science Foundation, the U.S. Department of Energy, the National Aeronautics and Space Administration, the Japanese Monbukagakusho, the Max Planck Society, and the Higher Education Funding Council for England.


\clearpage

\end{document}